\documentclass[aps,prd,reprint,groupedaddress]{revtex4-1}
\usepackage{hyperref}
\usepackage{graphicx}
\usepackage{color}
\usepackage{amsmath}
\usepackage{amssymb}
\usepackage{lipsum}
\hypersetup{
	pdfnewwindow=true,      
	colorlinks=true,       
	linkcolor=blue,          
	citecolor=blue,        
	filecolor=blue,      
	urlcolor=blue           
}

\newcommand{\mnras}{MNRAS}
\newcommand{\apjl}{ApJ Lett.}

\begin{document}
	
\title{How would a nearby kilonova look on camera?}

\author{Nihar Gupte}
\affiliation{Department of Physics, University of Florida, Gainesville, FL 32611, USA}
\author{Imre Bartos}
\affiliation{Department of Physics, University of Florida, Gainesville, FL 32611, USA}
	
\begin{abstract}
Kilonovae are cosmic optical flashes produced in the aftermath of the merger of two neutron stars. While the typical radiant flux of a kilonova can be as high as $10^{34}$\,W, they typically occur at cosmological distances, requiring meter-class or larger telescopes for their observation. Here we explore how a kilonova would look like from Earth if it occurred in the Solar System's backyard, 1000 light years from Earth. This is a small distance on cosmological scales, with only one neutron-star merger expected to occur this close every 100,000,000 years. While humans will likely only see kilonovae at much greater distances, showing how such a nearby event would look on camera can help visualize these events, and demonstrate their unique spectral evolution.
\end{abstract}
	
\maketitle
	
\section{Introduction}
	
Neutron stars are the ultracompact remnants of massive stars whose core collapsed under its own gravitational weight. They are typically 1-2 times the mass of our Sun, compressed in to a sphere of around 12\,km radius. 

Occasionally, two neutron stars collide in the Universe. This is usually due to the emission of gravitational waves that bring two neutron stars that orbit each other closer to each other until they merge. 

During this merger, the neutron stars shed some of their matter due to tidal and other effects. The resulting neutron-rich outflow creates heavy elements that undergo radioactive decay, and produce an energetic, week-long optical emission called {\it kilonova} \cite{1998ApJ...507L..59L,Kulkarni:895308,2005ApJ...634.1202R,2010MNRAS.406.2650M,2013ApJ...775...18B}. 

Kilonovae are extremely bright, having a peak radiant flux of around $10^{34}$\,W within hours after the neutron-star merger, which then gradually decays over a period of a week. The emission spectrum is initially blue/white that gradually moves towards red due to the decreasing temperature of the radiating gas.

Neutron-star mergers typically occur at large distances from Earth, and therefore kilonovae are also typically seen from far away; the first detected kilonova was about three billion light years away from us \cite{2013Natur.500..547T}. More recently, the LIGO and Virgo gravitational-wave observatories discovered a neutron-star merger, named GW170817, which also produced a detectable kilonova, at a distance of 130 million light years \cite{PhysRevLett.119.161101,Abbott_2017}. 

Over longer periods of time, however, mergers will occasionally happen at much closer distances. In particular, the composition of heavy elements in the early Solar System indicate that there has been a neutron-star merger at only 1000 light years from the pre-Solar nebula--the gas cloud that eventually formed the Solar System, about 100 million years before the Solar System was formed \cite{BartosMarkaNature}.

In this paper we consider this latter scenario and examine how a kilonova would look like from Earth, in particular on camera, from a distance of 1000 light years. At this distance, a kilonova would outshine the entire night sky, and could also be visible during daytime. Such a bright emission appears as an extended source in an image due to {\it flare} and {\it bloom} (two components of lens flare), which visualize brightness beyond the limited dynamic range of the camera for small light sources.
	
\section{Method}
\label{sec:method}

In order to reproduce the visibility of a nearby kilonova, we rely on the measured spectrum of the neutron-star merger GW170817, which is the only event so far with detailed spectral observations.

To generate an image of the nearby kilonova, we execute the following steps:	
\begin{itemize}
\setlength{\itemsep}{3pt}
\item Obtain kilonova spectrum.
\item Convert kilonova spectrum into RGB colors. 
\item Determine the perceived brightness.
\item Compute the effect of the lens flare.
\item Add the desired background.
\end{itemize}
	
\begin{figure}[h!]
		\label{Spectra}
		\centering
		\includegraphics[width=0.5\textwidth]{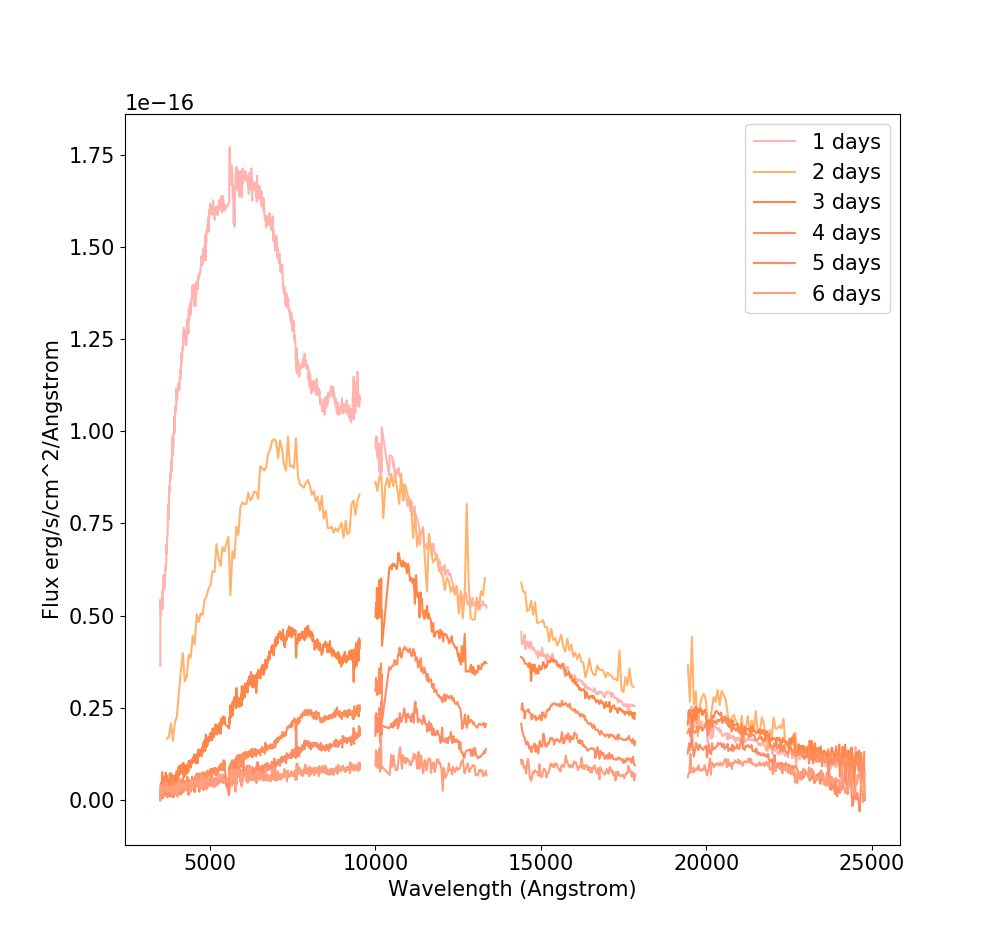}
		\caption{Spectrum data of GW170817 obtained from the Open Kilonova Catalog. The color of the line is the actual color of the merger at that time.}
	\end{figure}

First, we acquired the time-dependent spectrum of GW170817 from the "Open Kilonova Catalog" (https://kilonova.space/kne/GW170817/). The obtained spectrum is shown in Figure \ref{Spectra} at different times after the merger. 

Second, we converted the time-dependent spectrum to RGB colors visible to the human eye. Human vision is not directly linked to the spectrum. There are three different types of cones in the human eye with different sensitivities to wavelengths. The peaks for these three types occur between 420-440nm, 530-540nm and 560-580nm for the short, middle, long cones, respectively. 

\begin{figure}[h]
\label{cmfs}
\centering
\includegraphics[width=0.5\textwidth]{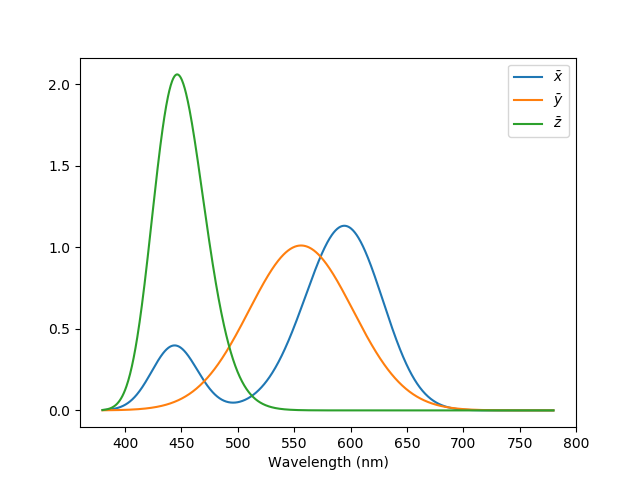}
\caption{Analytical approximations of the CIE 1931 color matching functions obtained from \cite{wyman2013simple}.}
\end{figure}

The response in the eye to brightness is non-linear. It depends on how dim the object is (using scotopic versus photopic functions). Here, we find the colors using the CIE "Standard Observer" by using the CIE color matching functions $\bar{x}, \bar{y}, \bar{z}$ \cite{cie1932commission}. We adopt the photopic color matching functions. Analytic approximations for these functions can be obtained from \cite{wyman2013simple}. A plot of these color matching functions is shown in Figure \ref{cmfs}. First we convolve the spectrum with the color matching functions and find their corresponding tristimulus values $X$, $Y$ and $Z$. This is done using the following integrals:
\begin{equation} \label{TristimulusX}
X = \int_{\lambda_{min}}^{\lambda_{max}} L(\lambda) \,\bar{x}(\lambda)\,d\lambda
\end{equation}
\begin{equation} \label{TristimulusY}
Y = \int_{\lambda_{min}}^{\lambda_{max}} L(\lambda) \,\bar{y}(\lambda)\,d\lambda
\end{equation}
\begin{equation} \label{TristimulusZ}
Z = \int_{\lambda_{min}}^{\lambda_{max}} L(\lambda) \,\bar{z}(\lambda)\,d\lambda
\end{equation}
Here $L(\lambda)$ is the spectral radiance and $\lambda$ is the wavelength measured in nanometers. These tristimulus values are the color values in the CIE color space. To convert them into RGB values we use a matrix transformation. Each display is different so distinct matrices need to be used for different displays. For this paper the HDTV XYZ to RGB matrix will be used:
	\begin{equation}
	\begin{bmatrix}
	R \\
	G \\
	B
	\end{bmatrix}
	=
	\begin{bmatrix}
	6.20584986 & -1.7174614 & -1.04788582 \\
	-2.71554014 & 5.51336937 & 0.09687197 \\
	0.19384968 & -0.39357359 & 2.9841102
	\end{bmatrix}
	\begin{bmatrix}
	X\\
	Y\\
	Z 
	\end{bmatrix}
	\end{equation}
	
\begin{figure}[h]
\label{RGBInterpolated}
\includegraphics[width=0.5\textwidth]{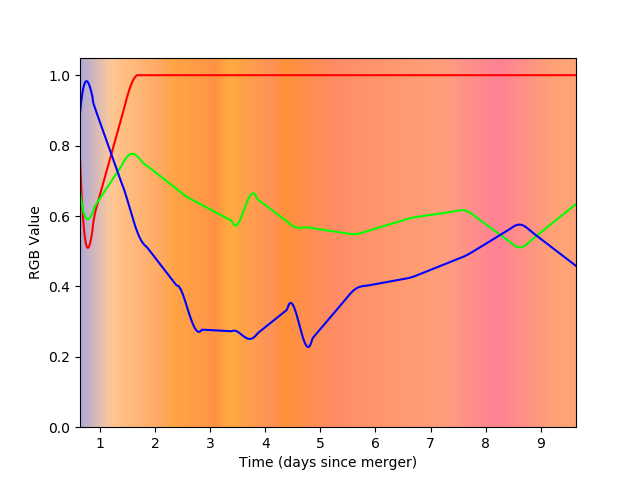}
\caption{Interpolated time dependent RGB values of the merger. The x-axis is time since the merger and y-axis is the RGB value normalized to 1.}
\centering
\end{figure}

Now that the RGB values are found as a function of time, we interpolate and smooth them. We use a cubic spline interpolation. We apply a Savitzy-Golay filter for smoothing \cite{savitzky1964smoothing}. Figure \ref{RGBInterpolated} shows the interpolated RGB values as a function of time. 
	
Next, we find the photometry from the spectrum. The most convenient band to measure the photometry is the Johnson-Cousin's "V" (visual) band \cite{bessell2005standard}. We take the spectrum of the merger and multiply it by the normalized V band function. We then integrate over the wavelength:
\begin{equation}
F = \int_{\lambda_{min}}^{\lambda_{max}} L(\lambda) V(\lambda) d\lambda,
\end{equation}
where $F$ is flux and $V(\lambda)$ is the Johnson radiometric response function in the visual band. We then rescale this flux to the desired distance, in this case 1000 light years.
	
Next, we create the merger. We employ the {\it lens flare} effect using Natron, a video editing software. A custom plugin from NatronVFX is used to create a lens flare. We change several features from the plugin in order to model the kilonova accurately. For example, we remove the "Anamorphic spots 3" node. Most prominently however, we use a point density function from \cite{spencer1995physically} (see equations 2 and 5) to model the "flare" and "bloom" of the merger accurately. To change the "flare" and "bloom" values into RGB we use a camera response function. The camera response function used here is for a Kodak DCS460 digital camera with a shutter speed of 1 second (see Equation 2 in \cite{debevec2008recovering}).

    \begin{figure}[h]
        \label{NodeGraph}
        \centering
        \includegraphics[width=0.47\textwidth]{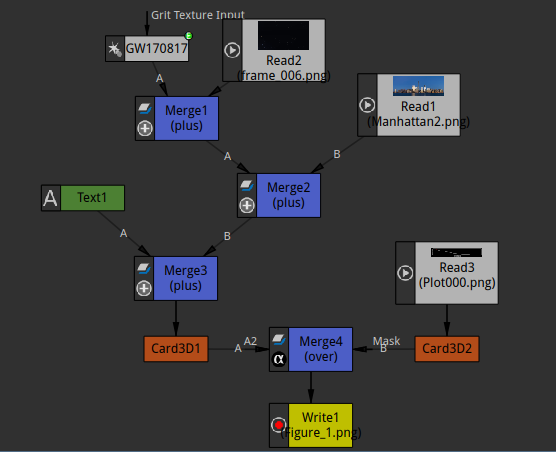}
        \caption{Natron Node graph of the video creation. For static backgrounds, the Read2 node will be deleted. The Card3D nodes are included to shift the graph with the background.}
    \end{figure}
	
The next step is the background. If a static background is desired, we only need to place the merger in a suitable position on the background. For a non-static background, one must take multiple background images. This can be done using various planetarium software, for example Stellarium. Stellarium allows setting various parameters including latitude, longitude, atmospheric effects, elevation etc. It also allows the setting of markers based on right ascension and declination or altitude and azimuth coordinates. Using this we can track the merger on the night sky. For example we show in Table \ref{table:params} the coordinates used for GW170817 over the Manhattan skyline (see Figure \ref{Manhattan}).
    
Now that we have taken background images we can place the merger at the appropriate location in the background image. This is accomplished by tracking a marker that is placed in Stellarium and placing the merger on the marker.

	\begin{center}
	\label{table:params}
		\begin{tabular}{|c | c|} 
			\hline
			Start Date & 08/18/2017 \\ 
			\hline
			End Date & 08/27/2017	\\
			\hline
			Right Ascension & 13h 9m 48.08s \\
			\hline
			Declination &  $-23^\circ 22^\prime 53.3^{\prime\prime}$\\
			\hline
			Latitude &  S $90^\circ 0^\prime 0.00^{\prime\prime}$\\
			\hline
			Longitude & W $139^\circ 15^\prime 59.99^{\prime\prime}$\\ [1ex] 
			\hline
			Elevation & 2835m \\
			\hline
		\end{tabular}
	\end{center}

After the merger has been created the last step is to combine the background images and the kilonova.  Using Natron we combine the 3 objects to create frames (see Figure \ref{NodeGraph} for a Node Graph). Then ffmpeg is used to combine these frames into a video.	
	
\section{Results}
	
\begin{figure*}
\label{Manhattan}
\centering
\includegraphics[width=\textwidth]{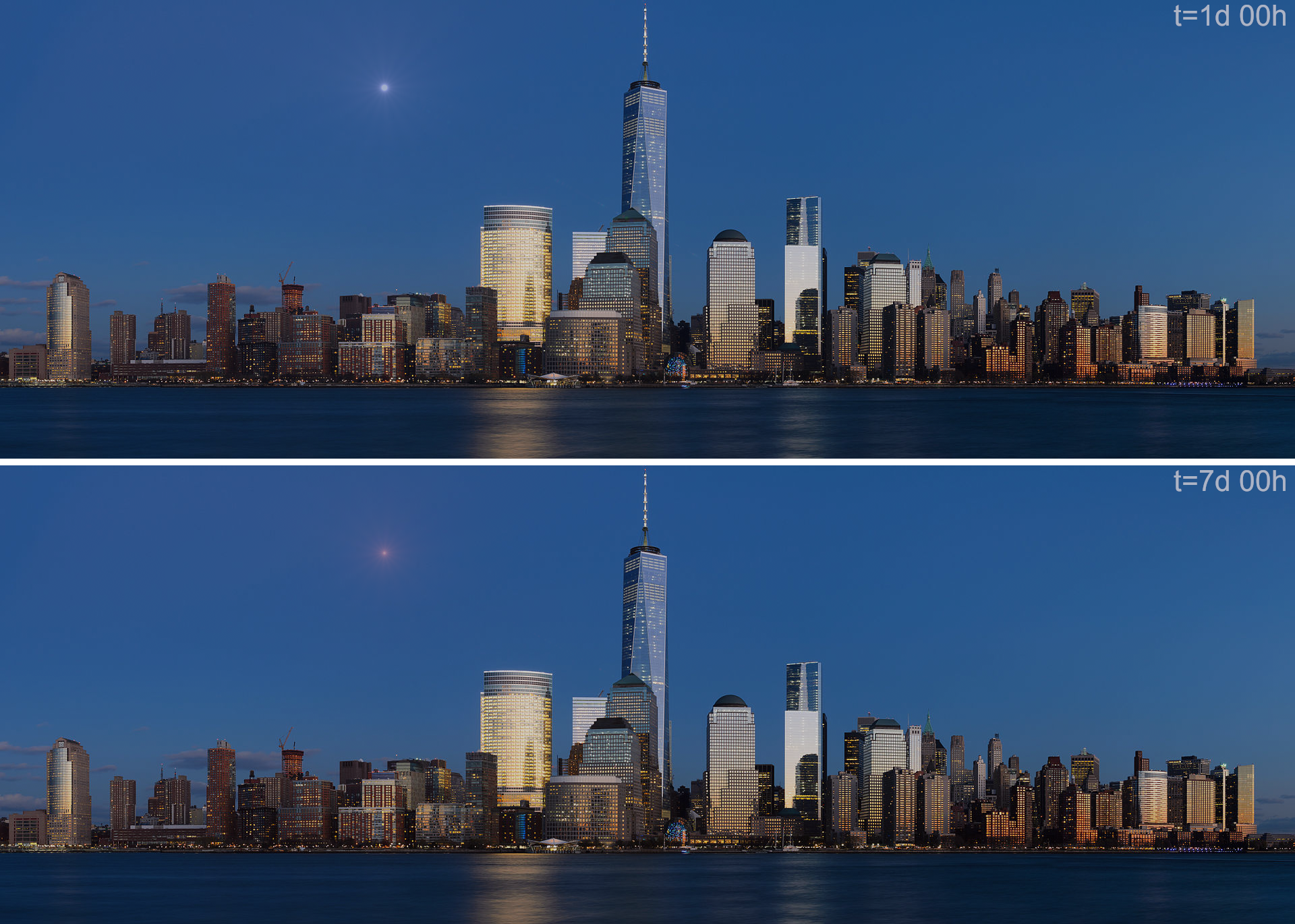}
\caption{Simulated images of a kilonova at 1000 light years from Earth viewed on camera behind the Manhattan skyline. Image at 1 day (top) and 7 days (bottom) after the neutron-star merger are shown. Background image credit: King of Hearts / Wikimedia Commons / CC-BY-SA-3.0 }
\end{figure*}

\begin{figure*}
\label{Perth}
\centering
\includegraphics[width=1\textwidth]{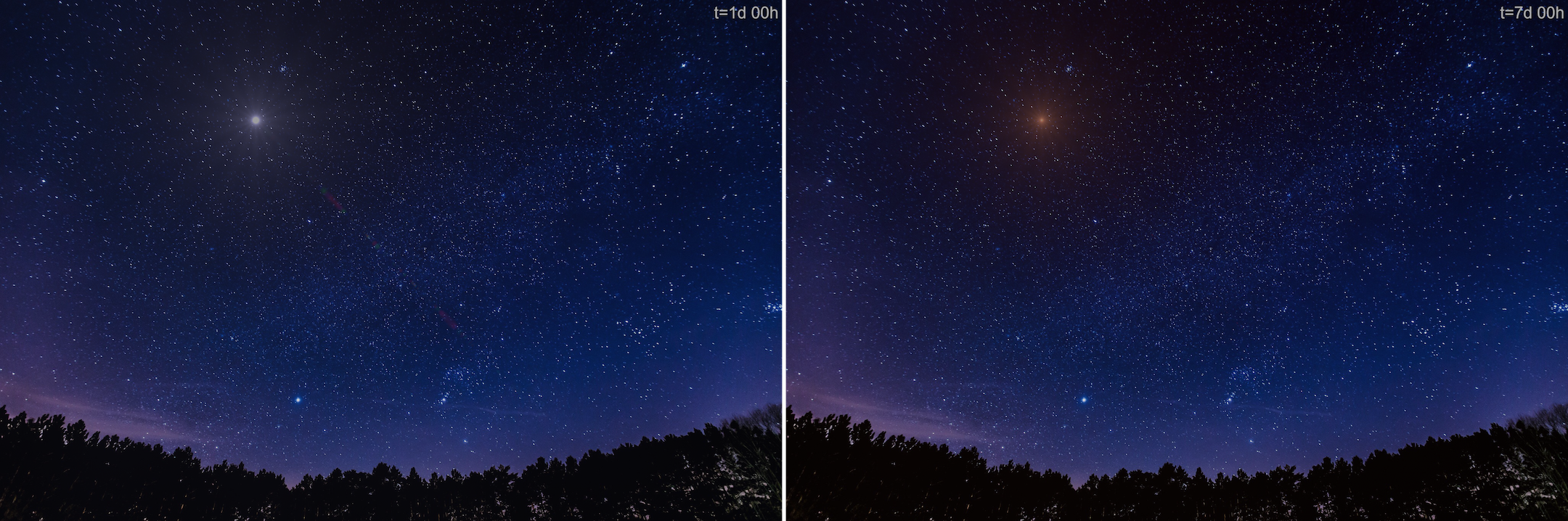}
\caption{Simulated images of a kilonova at 1000 light years from Earth viewed on camera during nighttime at Perth Australia. Image at 1 day (top) and 7 days (bottom) after the neutron-star merger are shown. Image credit: Pixelbay. }
\end{figure*}

\begin{figure*}
\label{Liberty}
\centering
\includegraphics[width=\textwidth]{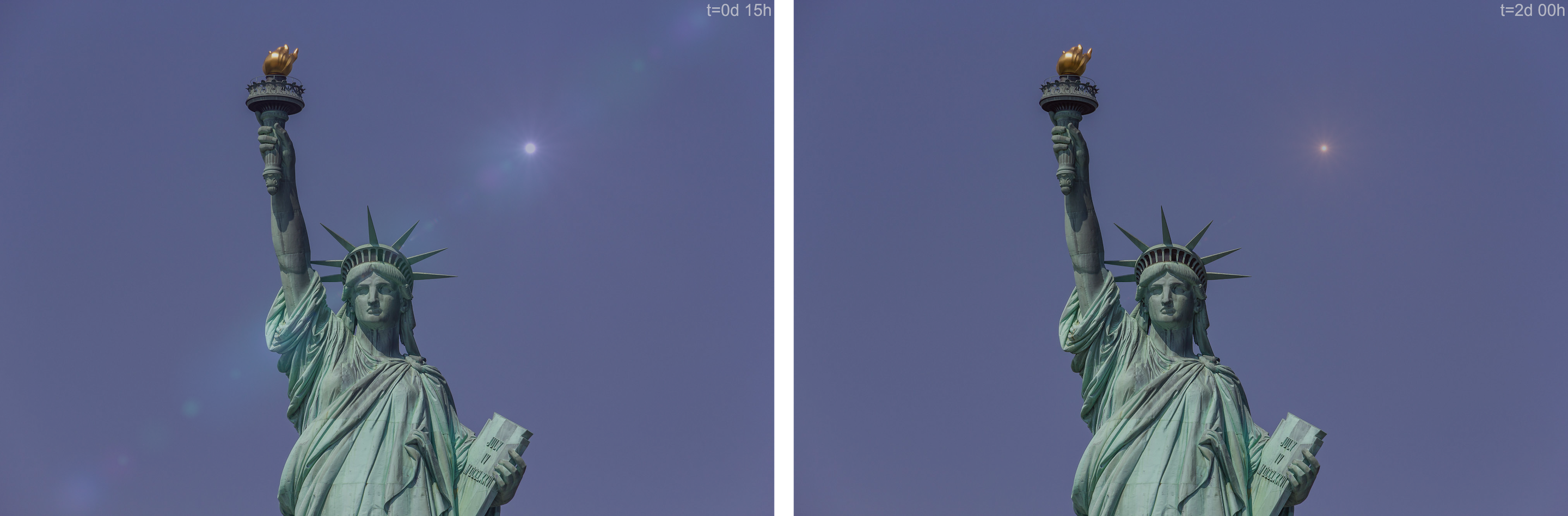}
\caption{Simulated images of a kilonova at 1000 light years from Earth viewed on camera behind the Statue of Liberty. Image at 15 hours (top) and 7 days (bottom) after the neutron-star merger are shown. Background image credit: Max Touhey.}
\end{figure*}

\begin{figure*}
\label{CenturyTower}
\centering
\includegraphics[width=0.9\textwidth]{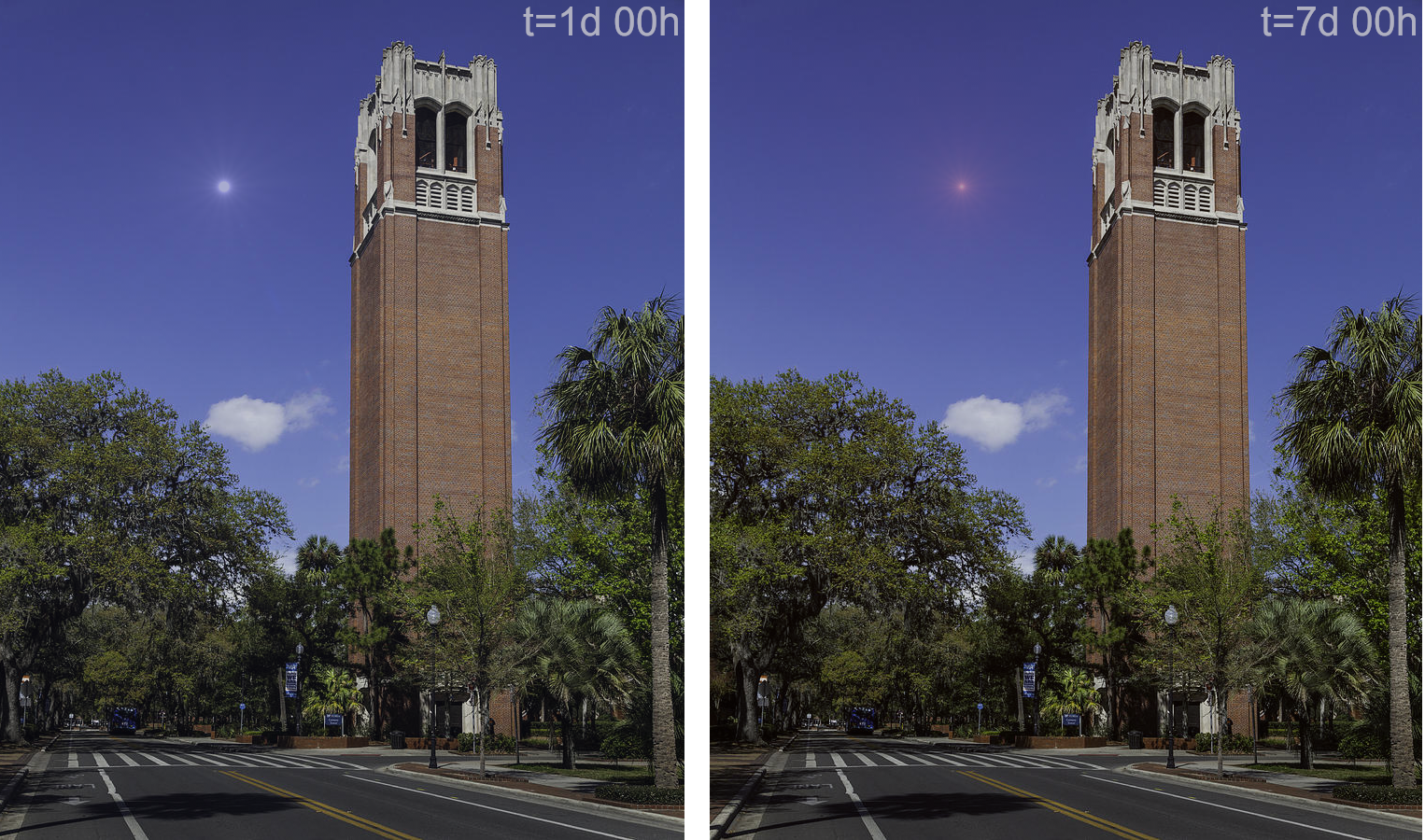}
\caption{Simulated images of a kilonova at 1000 light years from Earth viewed on camera behind the Century Tower at the University of Florida. Image at 1 day (top) and 7 days (bottom) after the neutron-star merger are shown. Background image credit: Lynn Palmer.}
\end{figure*}

\begin{figure*}
\label{FortKnox}
\centering
\includegraphics[width=0.9\textwidth]{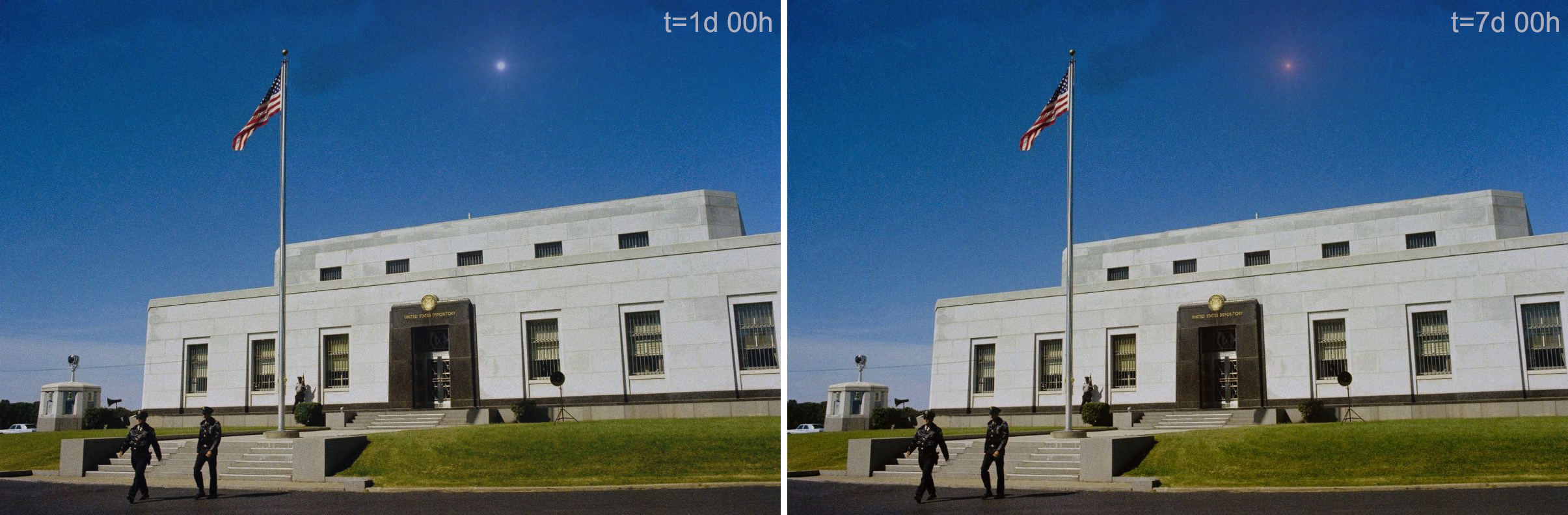}
\caption{Simulated images of a kilonova at 1000 light years from Earth viewed on camera behind Fort Knox in Kentucky. Image at 1 day (top) and 7 days (bottom) after the neutron-star merger are shown. Background image credit: Bettmann/Getty Images}
\end{figure*}

We selected several background images to demonstrate how a nearby kilonova would look like. These include a skyline view of Manhattan (Figure \ref{Manhattan}), a nighttime image (Figure \ref{Perth}), the Statute of Liberty (Figure \ref{Liberty}), the Century Tower at the University of Florida (Figure \ref{CenturyTower}), and Fort Knox (Figure \ref{FortKnox}).	
	
\section{Conclusion}

We described how to reproduce the image of a nearby kilonova on camera. We add a lens flare (both flare and bloom) effect on images and indicate the brightness of a point source beyond the camera's dynamic range.

\bibliography{Refs}
\end{document}